# Nonoptimal Component Placement, but Short Processing Paths, due to Long-Distance Projections in Neural Systems

Marcus Kaiser[1,2,3*], Claus C. Hilgetag[3,4]

1 School of Computing Science, University of Newcastle, Newcastle upon Tyne, United Kingdom, 2 Institute of Neuroscience, University of Newcastle, Newcastle upon Tyne, United Kingdom, 3 International University Bremen, School of Engineering and Science, Bremen, Germany, 4 Boston University, Sargent College, Department of Health Sciences, Boston, Massachusetts, United States of America

It has been suggested that neural systems across several scales of organization show optimal component placement, in which any spatial rearrangement of the components would lead to an increase of total wiring. Using extensive connectivity datasets for diverse neural networks combined with spatial coordinates for network nodes, we applied an optimization algorithm to the network layouts, in order to search for wire-saving component rearrangements. We found that optimized component rearrangements could substantially reduce total wiring length in all tested neural networks. Specifically, total wiring among 95 primate (Macaque) cortical areas could be decreased by 32%, and wiring of neuronal networks in the nematode *Caenorhabditis elegans* could be reduced by 48% on the global level, and by 49% for neurons within frontal ganglia. Wiring length reductions were possible due to the existence of long-distance projections in neural networks. We explored the role of these projections by comparing the original networks with minimally rewired networks of the same size, which possessed only the shortest possible connections. In the minimally rewired networks, the number of processing steps along the shortest paths between components was significantly increased compared to the original networks. Additional benchmark comparisons also indicated that neural networks are more similar to network layouts that minimize the length of processing paths, rather than wiring length. These findings suggest that neural systems are not exclusively optimized for minimal global wiring, but for a variety of factors including the minimization of processing steps.

## Introduction

The organization of neural systems is shaped by multiple constraints, ranging from limits placed by physical and chemical laws to diverse functional requirements. In particular, it is of interest to identify factors influencing the layout of neural connectivity networks. One prominent idea is that the establishment and maintenance of neural connections carry a significant metabolic cost that should be reduced wherever possible [1,2]. As a consequence, wiring length should be globally minimized in neural systems. This requirement places strong constraints on the design of neural networks across different levels of organization [3–8].

A trend toward wiring minimization is apparent in the distributions of projection lengths for various neural systems, which show that most neuronal projections are short [9–12]. However, wiring length distributions also indicate a significant number of longer-distance projections, which are not formed between immediate neighbours in the network. Such projections may be required for functional reasons, so that a strictly minimal wiring of networks, using only local projections, is not attainable.

Alternatively, it has been suggested that wiring length reductions in neural systems are achieved not by minimal rewiring of projections within the networks, but by suitable spatial arrangement of the components. Under these circumstances, the connectivity patterns of neurons or regions remain unchanged, maintaining their structural and functional connectivity, but the layout of components is perfected such that it leads to the most economical wiring. In the sense of this "component placement optimization" (CPO [3]), any rearrangement of the position of neural components, while keeping their connections unchanged, would lead to an increase of total wiring length in the network. CPO has been reported for neural networks at different levels of organization, such as interconnected ganglia in the nematode *Caenorhabditis elegans* [3] and cerebral cortical areas in rat, cat, and monkey brains [3,5,6,13]. Moreover, optimal component placement was suggested to apply to the arrangement of human cranial nerves [3] as well as the layout of cortical maps [14], and mechanisms yielding optimal placement were proposed [15,16].

We revisited the concept of CPO and investigated the layout of two representative neural networks in metric space, analyzing three-dimensional spatial positions of connected

Abbreviations: CPO, component placement optimization

* To whom correspondence should be addressed. E-mail: M.Kaiser@ncl.ac.uk

## Synopsis

What constraints shape the organization and spatial layout of neural networks? One influential idea in theoretical neuroscience has been that the overall wiring of neural networks should be as short as possible. Wire-saving could be achieved, for instance, through an optimal spatial arrangement of the connected network components. The authors evaluated this concept of component placement optimization in two representative systems, the neuronal network of the *Caenorhabditis elegans* worm and the long-range cortical connections of the primate brain. Contrary to previous results, they found many network layouts with substantially shorter total wiring than that of the original biological networks. This nonoptimal component placement arose from the existence of long-distance connections in the networks. Such connections may come at a developmental and metabolic cost; however, as the analyses reported in this article show, they also help to reduce the number of signal processing steps across the networks. Therefore, the organization of neural networks is shaped by trade-offs from multiple constraints, among them total wiring length and the average number of processing steps.

cortical areas in the primate (Macaque) brain and two-dimensional positions of individual neurons in the neuronal network of *C. elegans*. Due to the more limited availability of data in the past, previous analyses of these networks [3,6] included less detail and fewer components and were based on neighbourhood (adjacency) relationships, rather than on metric spatial coordinates of the components. We investigated component rearrangements in the networks with the help of a simulated annealing search approach and found that, remarkably, the extended and refined datasets possessed no optimal component placement. Instead, the spatial rearrangement of network components could lead to substantial wire saving, due to a considerable number of long-distance projections in the networks.

While the role of long-range connections, or network shortcuts, has been previously explored in topological analyses of neural connectivity [7,11], the specific role of long-distance projections in the spatial layout of neural systems is still uncertain. In order to address this question and to explore alternative constraints on the organization of neural connectivity, we also compared the biological neural networks with different benchmark networks of the same size in which connections were rewired minimally or randomly, or same-size networks optimized for minimum and maximum wiring length and average path lengths, respectively. The comparisons demonstrated that biological neural networks feature shorter average path lengths than networks lacking long-distance connections. Moreover, the average path lengths of neural networks, corresponding to the average number of processing steps, were close to path lengths in networks optimized for minimal paths.

## Results

### Overview

First, we derived the wiring length distribution of the primate and *C. elegans* network (Figure 1A–1C). Second, we used an optimization approach based on simulated annealing to search for wire-saving component rearrangements in the two networks (see Figure S1). For both networks, alternative component rearrangements existed that resulted in substantially reduced total wiring (Figure 1D–1F). The rearranged networks showed a decrease in the number of long-distance connections (Figures 1G–1I, 2, and 3). Third, we explored the role of long-distance projections, by comparing the original neural networks with minimally rewired networks not possessing long-distance connections. The relative comparisons, carried out against the background of a variety of benchmark networks (Figures 4 and 5), demonstrated that long-distance connections may confer adaptive benefits, particularly by reducing the number of intermediate processing steps in neural networks.

### Wiring Length Distribution

In the primate brain and in the neuronal network of *C. elegans*, the reach of connections among components quickly decays with distance (Figure 1A–1C). Nonetheless, some connections are not formed between immediate neighbours and extend over a considerable distance. For example, more than 10% of the primate cortical projections connect components that are separated by more than 40 mm (more than half of the maximally possible spatial distance between components of 69 mm; Figure 1A). A similar distribution emerged for the local connectivity in *C. elegans* (Figure 1B). For the global *C. elegans* network, some connections were almost as long as the entire organism (Figure 1C). Thus, biological neural networks are not strictly minimally wired in the sense that only the shortest possible connections are established ([11]; also see section "Minimally rewired networks", below). However, their components may be spatially arranged in such a way that the overall wiring is minimal, given the specific connectivity of the network. This hypothesis is tested in the next section.

### Reduction of Total Wiring Length by Component Rearrangement

We tested the concept of CPO, which states that any spatial rearrangement of neural network components leads only to an increase, not a decrease, of total network wiring. As an exhaustive search of all possible alternative node arrangements was not feasible for the given large-scale networks, we used a simulated annealing algorithm to specifically search for overall wiring reductions in spatially permuted component layouts (see Materials and Methods for details). Briefly, at each step three nodes were cyclically rearranged, and testing explored whether the new solution led to a shorter total wiring length. The algorithm quickly converged and led to an approximate solution for the optimal wiring layout of the components (Figure S1). Note that this approach modified only spatial node positions, whereas the network topology (the specific afferent and efferent connections of each node) remained unchanged.

**Macaque: Cortical area rearrangements.** Area rearrangement by simulated annealing reduced the total wiring length of the primate cortical network by up to 32% (Figure 1D). In the wire-saving new arrangement, the overall number of long-distance connections was reduced (Figure 1G). This reduction resulted from placing areas with many projections (e.g., area V1) closer to the areas to which they are mainly connected. Moreover, areas possessing fewer connections were moved to the spatial periphery of the rearranged network (Figure 2).

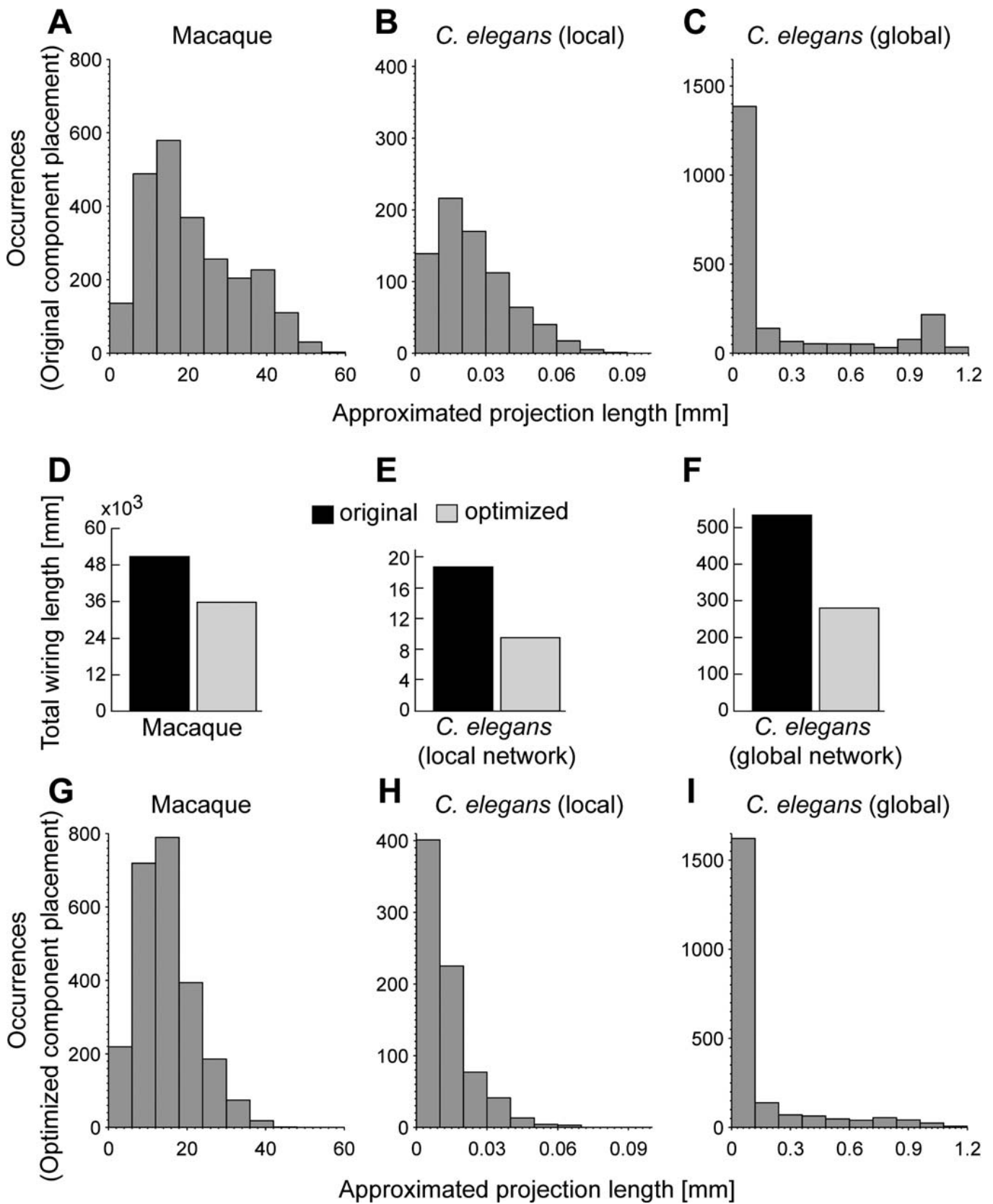

**Figure 1.** Projection Length Distribution and Total Wiring Length for Original and Rearranged Neural Networks

(A–C) Approximated projection length distribution in neural networks. Macaque monkey cortical connectivity network with 95 areas and 2,402 projections (A). Local distribution of connections within rostral ganglia of *C. elegans* with 131 neurons and 764 projections (B). Global *C. elegans* neural network with 277 neurons and 2,105 connections (C).
(D–F) Reduction in total wiring length by rearranged layouts yielded by simulated annealing for Macaque cortical network (D), *C. elegans* local network (neurons within rostral ganglia) (E), and global *C. elegans* network (F).
(G–I) Approximated projection length distribution in neural networks with optimized component placement. Macaque monkey cortical connectivity network (G). Local distribution of connections within rostral ganglia of *C. elegans* (H). Global *C. elegans* neural network (I). For all optimized networks, the number of long distance connections is reduced compared to the original length distribution in (A)–(C).
DOI: 10.1371/journal.pcbi.0020095.g001

**Macaque monkey: Influence of cortical area sizes.** We considered two potential confounds of the spatial rearrangement analysis. First, swapping areas with different area sizes might also affect area positions. For example, exchanging a small cortical area such as the lateral intraparietal area (about 50 mm$^2$ [17]) with area V1 or V2 (size about 1,200 mm$^2$ [17]) would also result in shifting the positions of other cortical areas, and ultimately in changing total wiring length. Therefore, in a modified approach, we limited spatial permutations to areas whose surface sizes did not differ by more than 5%. This constraint substantially restricted the number of permissible rearrangements, as only 4% of all possible swaps met the same-size criterion. Despite this restriction, total wiring length in the cortical network could still be reduced by 12.5%.

**Macaque monkey: Role of white matter volume.** As a second potential confound, the rearrangement analysis investigated metric projection distances between cortical areas, but not white matter volume. However, two fibre tracts of the same metric length might differ in the actual number of axons that form the projections. Therefore, fibre tract diameter or volume would constitute a better measure of total axonal wiring. Unfortunately, no systematically collated metric information is available for the diameters of corticocortical fibres. Therefore, white matter volume of specific projections cannot be calculated in a straightforward way. An approximation used in earlier studies of CPO [6] was to employ connection strength or density as an estimate of white matter volume. One of the sources for the present cortical network data [18] reported connection strengths for projections among 18 primate prefrontal areas as ordinal values: 0 (absent or unknown), 1 (light), 2 (moderate), and 3 (heavy). We used this cortical subnetwork of 18 areas to explore the role of connection strength in wire-saving component arrangements. Total wiring volume was calculated by multiplying the distance between connected nodes by the square of the connection strength of the respective projection, since the cross section of a fibre is proportional to the square of its diameter (therefore, cross-section areas were 1, 4, and 9, respectively). Thus, for two connections with the same length, but respective connection strengths 1 and 3, the connection with strength 3 was assigned a white matter volume nine times as large as the connection with strength 1. With these measures in the simulated annealing search algorithm, the total wiring volume of projections among prefrontal areas could still be reduced by 16%. By comparison, the approach of optimizing total wiring length without accounting for connection strength led to a reduction of 10% in the wiring among these 18 prefrontal areas. Therefore, considering total wiring volume instead of total wiring length did not appear to alter the principal conclusion of suboptimal component placement in cortical networks. However, further investigations, using improved information about fibre bundle diameter as well as larger datasets, need to be conducted.

***C. elegans*: Neuronal network rearrangements.** For the global neural network of *C. elegans,* which includes many long-distance connections (Figure 1C), the component rearrangement algorithm produced a maximum wiring reduction of 48%. Meanwhile, rearrangement of neurons in the rostral ganglia alone reduced total wiring length by 49% (Figure 1E and 1F). In line with the lowered total wiring length, the number of long-distance connections also decreased (Figure 1H and 1I), as described for the primate cortical network above. During optimization by simulated annealing, connections aligned with the longitudinal (horizontal) axis were rearranged along the shorter vertical axis for the local *C. elegans* network within rostral ganglia (Figure 3A and 3B). Similarly, rearranging neuron positions in the global *C. elegans* network (Figure 3C and 3D) reduced the number of long-distance connections running along the longitudinal axis. These findings also remained valid when variations in connection site, of the third spatial coordinate, or in synaptic type were taken into account (see Materials and Methods, "Control calculations for *C. elegans* network").

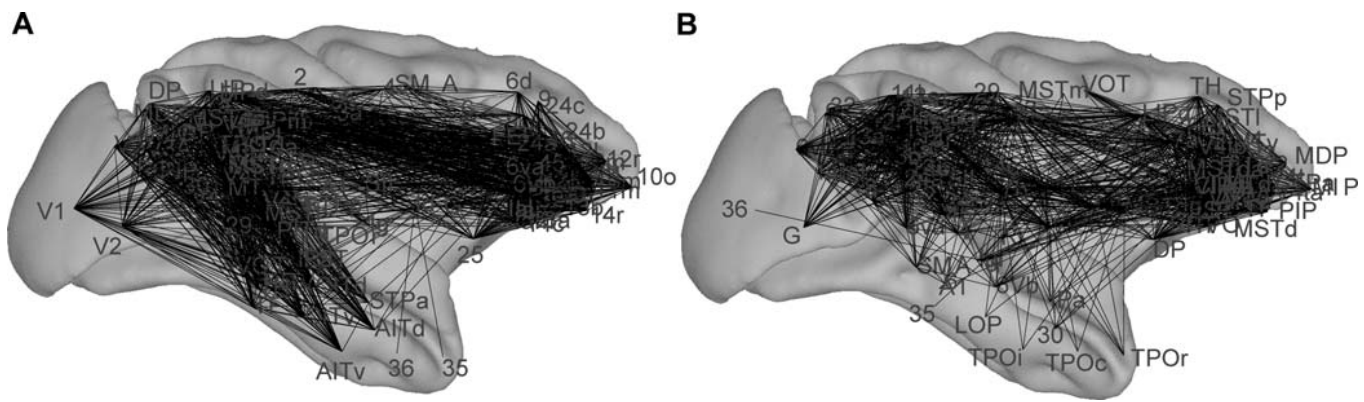

**Figure 2.** Original and Optimally Rearranged Macaque Cortical Networks
(A) Original placement of 95 cortical areas.
(B) Network layout after evolutionary rearrangement of areas to minimize total wiring.
A larger version of this figure is available at http://www.biological-networks.org.
DOI: 10.1371/journal.pcbi.0020095.g002

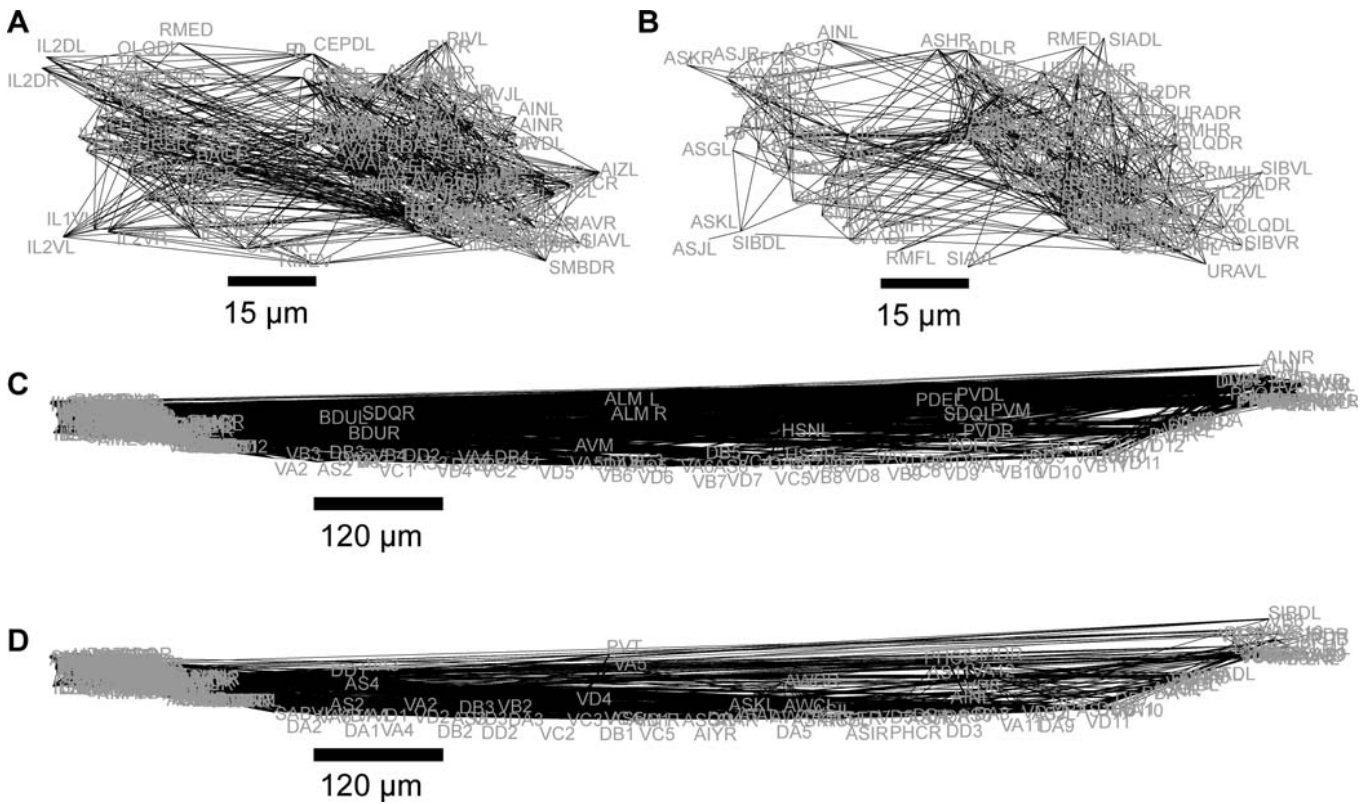

**Figure 3.** Original and Optimally Rearranged Layouts of Local and Global Neural Networks of *C. elegans*
(A) Original placement of neurons within rostral ganglia.
(B) Optimized wire-saving component placement of rostral ganglia neurons.
(C) Original layout of global *C. elegans* network (lateral view).
(D) Global *C. elegans* neuronal network, rearranged to minimize total network wiring.
A larger version of this figure is available at http://www.biological-networks.org.
DOI: 10.1371/journal.pcbi.0020095.g003

### Alternative Constraints on Network Organization

**Minimally rewired networks.** For both the primate cortical and the *C. elegans* network, rearrangement of components yielded shorter total wiring lengths by reducing the number of long-distance connections (Figures 1A–1C versus 1G–1I, 2, and 3). Thus, biological neural networks possess more long-distance connections than do networks with strictly optimized component placement. To explore the potential benefits of long-distance connections, we compared the original neural networks with minimally rewired networks in which component positions and numbers of connections remained unchanged, but in which neighbouring nodes were preferentially connected (see Materials and Methods, "Minimal rewiring of networks"). Thus, very few long-distance connections existed in these networks, which represented the extreme state of adaptation to global minimal wiring.

We characterized the original and rewired networks with several network indices (Table 1). For the minimally wired networks, the clustering coefficient [19]—which describes the average connectivity between neighbours of a node—was higher than in the original networks (77% versus 64% in the primate cortical network; 43% versus 17% in the global *C. elegans* neuronal network; and 51% versus 14% in the local *C. elegans* neuronal network). Moreover, minimally rewired networks showed a lower total wiring length, leading to a global wire reduction of up to 90% (Figure 4A). Also, the average metric path length of the shortest path between two components was shorter in minimally rewired networks (Figure 4B). However, minimally rewired networks had one notable structural disadvantage: They possessed significantly longer average path lengths, as measured by the average number of connection segments between any two components, than did the biological networks (Figure 4C). This result is intuitively plausible, since the rewired networks lacked the network shortcuts provided by long-distance projections, so that all paths had to be routed via local neighbourhoods. For example, while direct connections between the occipital and frontal lobe exist in the original primate cortical network, such connections were absent in the minimally rewired network. In this case, the path between both regions involved several short-distance connections.

**Relative comparison of constraints.** Although the total wiring length in all analyzed networks could be reduced by node rearrangement, it might be the case that the original node configuration is already very close to the optimum compared to alternative arrangements. Therefore, we compared the actual wiring length of the different neural networks on a relative scale, on which the optimized wire-saving spatial arrangement yielded by simulated annealing

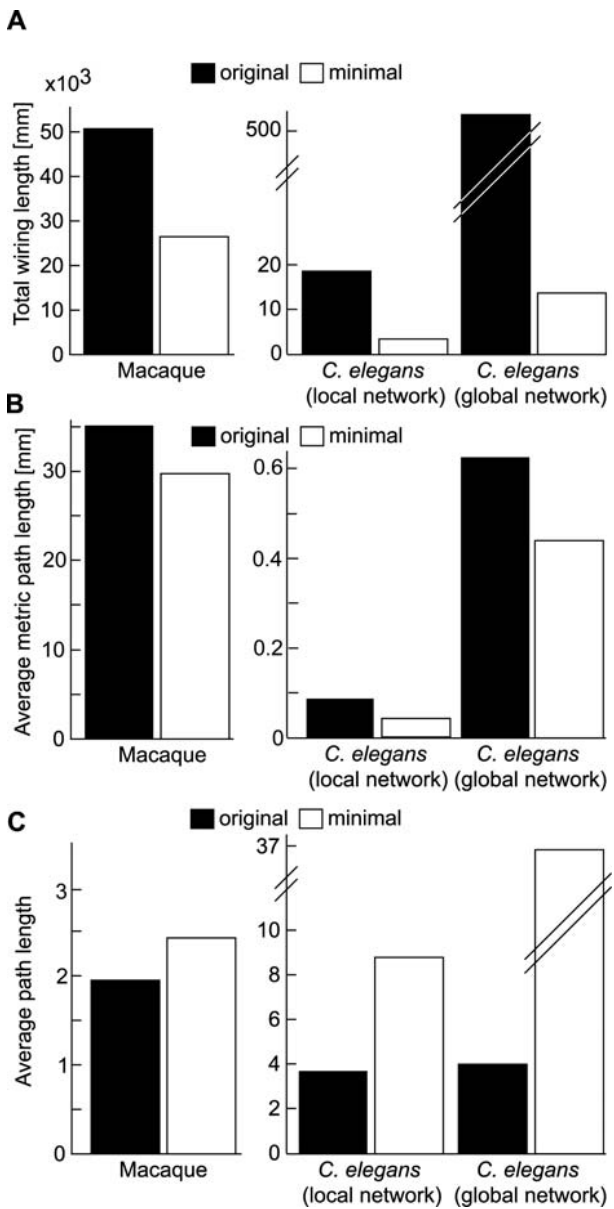

**Figure 4.** Network Properties of Original Cortical Networks and Minimally Rewired Networks of the Same Size Lacking Long-Distance Connections

Total wiring length (A) is substantially reduced in minimally rewired networks. Average metric length of the shortest path between any two nodes (B) is also lowered in the rewired networks. However, average path length (C), corresponding to the number of processing steps in the shortest path between components, is considerably smaller in the original than in the minimally rewired networks.
DOI: 10.1371/journal.pcbi.0020095.g004

represented the lower boundary (relative value "0"). The upper boundary ("1") was provided by an arrangement, also obtained by simulated annealing, in which the components were rearranged for maximum total wiring length. This relative comparison demonstrated that the actual wiring of neural networks was far from the optimal configuration (Figure 5A).

Moreover, an exhaustive search of all two-node and all three-node permutations in these networks also revealed a substantial proportion of alternative configurations with shorter total wiring: 29% of the two-node (and 17% of the

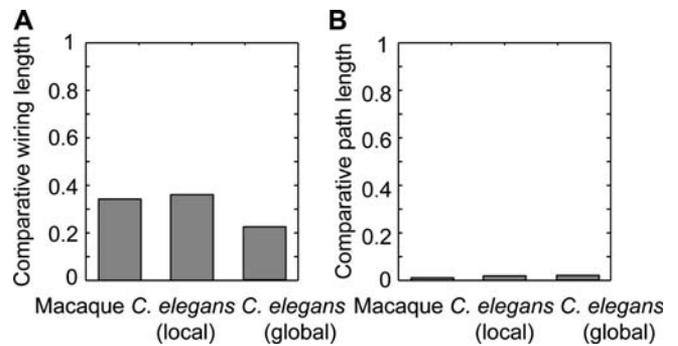

**Figure 5.** Wiring Arrangement of Neural Networks Compared to Minimum and Maximum Case Benchmark Networks

(A) Actual total wiring length relative to the minimum wiring length solution (value "0," yielded by simulated annealing of component positions) and to networks optimized for maximum total wiring length (value "1," also yielded by simulated annealing). The wiring of the different neural networks lies close to the middle between minimum and maximum case component arrangements.

(B) Average shortest path length (characteristic path length) in neural networks relative to networks optimized for minimum path length (value "0," yielded by simulated annealing of wiring organization) and maximum path length (value "1"). Actual path lengths in the neural networks are close to the lower bound of networks optimized for minimum paths.
DOI: 10.1371/journal.pcbi.0020095.g005

three-node) permutations in the primate network, 32% (26%) of permuted alternative configurations in the local *C. elegans* network, and 19% (13%) of the permutations in the global *C. elegans* network displayed shorter wiring than did the original configurations.

A different pattern emerged when the number of processing steps in the neural networks, as indicated by the average path length, was also compared on a relative scale. Here, the boundaries (relative values "0" or "1") were set by the path length of networks rewired for minimum or maximum average path lengths, respectively. Such networks were obtained by a simulated annealing search similar to that for optimal spatial rearrangements (see Materials and Methods for details). As shown by Figure 5B, the average path lengths of the cortical and *C. elegans* networks were close to those of same-size networks optimized for minimum path length.

As an additional benchmark, we also investigated randomly organized networks. The total wiring length for a random arrangement of nodes had relative values of $0.518 \pm 0.030$ (Macaque), $0.687 \pm 0.026$ (*C. elegans* global), and $0.544 \pm 0.028$ (*C. elegans* local). On the other hand, the random arrangement of edges resulted in relative values for average path length of $0.011 \pm 9.6 \times 10^{-17}$ (Macaque), $0.021 \pm 3.5 \times 10^{-17}$ (*C. elegans* global), and $0.018 \pm 3.0 \times 10^{-17}$ (*C. elegans* local).

## Discussion

### Summary

The distribution of projection lengths (Figure 1A–1C) and the absence of optimal component placement in diverse neural networks (Figure 1D–1F) suggest that wiring minimization may not be a predominant constraint on the design of neural networks at all levels. Instead, adding long-distance projections, and thereby reducing the number of processing steps across the system, might for some neural networks

Table 1. Network Measures for the Original Neural Systems as well as for Minimally Rewired Networks of the Same Size

| Neural System | Original Networks | | | | | Minimally Rewired Networks | | | |
|---|---|---|---|---|---|---|---|---|---|
| | TW | AMP | AP | CC | D | TW | AMP | AP | CC |
| Macaque monkey | 50,455 | 35.0 | 1.9 | 0.643 | 0.269 | 26,555 | 29.81 | 2.45 | 0.766 |
| C. elegans (global) | 532 | 0.625 | 4.0 | 0.167 | 0.028 | 13.5 | 0.4439 | 36.9 | 0.425 |
| C. elegans (local) | 18.6 | 0.0726 | 3.7 | 0.143 | 0.045 | 3.0 | 0.0436 | 8.79 | 0.511 |

The edge density of the minimally rewired networks is not shown, as it is identical to that of the original networks.
AMP, average metric path length (mm); AP, average path length; CC, clustering coefficient; D, edge density; TW, total wiring length (mm).
DOI: 10.1371/journal.pcbi.0020095.t001

outweigh the costs of establishing and maintaining additional fibre tracts. In the following sections we discuss how our analysis differed from previous studies that found evidence for CPO in neural networks, whether CPO might arise at some levels of neural organization but not others, and how CPO may be balanced with alternative demands on the organization of neural systems.

### Differences between the Present and Previous Studies

**Datasets.** Our findings, which are based on the analysis of extended and refined spatial representations of neural networks, are in apparent contrast to the widely reported optimal component placement in neural systems ([3,5,6,13], but see [20,21]). Previous analyses, however, were more limited in their approach. For example, earlier studies considered only a subset of 11 areas in the Macaque prefrontal cortex employing two-dimensional coordinates and adjacency relations as distance measure [6], or 17 Macaque and 15 cat visual areas, for which distance was also measured by adjacency relations [13]. Other investigations examined the adjacency arrangement of 11 entire C. elegans ganglia [3], but not the actual spatial positions of neurons within the ganglia nor the positions of all individual C. elegans neurons as analyzed here. Because of the smaller number of nodes in the previously analyzed datasets, fewer opportunities existed for wire-saving rearrangements than for the larger sets of 95 (primate) or 277 (C. elegans) components considered here. Moreover, adjacency relations provide more coarsely defined distance measures than did the continuous Euclidean distances used in the present study, also resulting in more limited degrees of freedom for wire reductions. Finally, previous datasets were analyzed as one-dimensional (C. elegans) or two-dimensional (primate) layouts, and the present consideration of additional dimensions expanded the search space for wire-saving component rearrangements.

**Analysis approaches.** The choice of the employed analytical techniques is related to the size of analyzed datasets. While it was possible to exhaustively test all possible component arrangements for the smaller datasets in previous studies, a stochastic optimization approach (simulated annealing) had to be applied for the large networks used in the current study. However, this approach also brought methodological advantages, as only small modifications in component placement were made at each step of the annealing algorithm. Moreover, the algorithm searched specifically for improvements in total wiring length. We note that an alternative approach, of randomly sampling different component rearrangements in the hope of identifying layouts with lower total wiring length, would have been unlikely to succeed. Scanning such random samples of primate area rearrangements, we found that all permutations had longer total wiring length compared with the original network, increasing wiring by at least 14%. This was due to the fact that more than 85 areas were placed at new positions in each of these permutations, and wire-saving effects of more limited exchanges involving only two to five nodes were hidden.

However, when we exhaustively searched all configurations that could be obtained from the original network arrangements by permutation of just two or three nodes (a large search space with up to 21,024,300 configurations that represents the practical limit of exhaustive analysis), we also found a high proportion (up to 32%) of permuted configurations with shorter wiring length. Therefore, the analysis revealed abundant alternative network arrangements with shorter wiring than in the original neural networks.

The current study used three-dimensional Euclidean distances between cortical areas as a metric measure of inter-area wiring length. This is an improvement over previous approaches in which adjacency relations on a two-dimensional cortical map (e.g., "neighbour," "next-neighbour-but-one") were used as ordinal estimates of wiring length. While the actual length of fibres in the cerebral network is still inaccessible due to practical limitations of current experimental techniques, recent work suggests that only a minority of corticocortical projections are strongly curved, whereas the majority of projections are straight or just mildly curved. For instance, only about 15% of the connections among prefrontal cortices in the Macaque monkey possess strongly curved trajectories, and dense fibres, in particular, tend to be completely straight [22]. Therefore, Euclidean distances may represent a reasonable approximation for the actual length of most projections.

**Component placement optimization may differ for different neural networks.** Given the more restricted focus of earlier studies finding CPO, it also appears possible that optimal placement exists at some levels of neural organization but not others. For instance, connectivity within the prefrontal subnetwork of the primate cortex may be dominated by connections to neighbouring areas; and significant numbers of long-range projections might only arise from the interconnections of prefrontal cortices with structures in other lobes of the brain. Indeed, an area rearrangement analysis of 11 prefrontal areas studied previously [6] confirmed optimal component placement within this restricted network (unpublished data). In addition, our rearrangement analysis of 18 prefrontal areas defined by three-dimensional positions identified a possible reduction of 10%, much lower than the 32% reduction obtained for the global network. Therefore, optimal component placement of the 11 prefrontal areas could be due to the smaller subset of areas involved as well as the greater number of their local interconnections. For the prefrontal network, for example, edge density was much higher than for the global network (67% compared to 27%). Similarly, optimal placement, based on adjacency of connected components, was found for comparably small datasets of closely linked areas in the cat and Macaque visual cortex [13].

Optimal component placement is also more likely to occur when competing design constraints are absent in specific neural populations, for example, when a reduction of the number of processing steps is impossible. One example is the parallel wiring between two directly connected regions where intermediate processing steps do not exist. Indeed, optimal component placement was found for the mapping of fibres between ocular dominance columns [14], or the vertical integration across cortical layers [8].

## CPO and Alternative Constraints on Neural Network Organization

Neural networks contain a substantial proportion of long-distance projections, many more than minimally rewired networks of the same size. Due to these far-reaching spatial shortcuts, minimal rewiring of the biological networks led to a global reduction in the amount of wiring and to a reduced metric length of the average shortest path between components. Moreover, minimally rewired networks showed increased clustering of connections within local neighbourhoods (Table 1). Such features are potentially beneficial for the organization of neural networks, resulting in economical wiring as well as greater local integration of network nodes.

However, minimal network rewiring also resulted in significantly increased average path length (corresponding to the number of connection steps in the shortest pathways) between the components (Figure 4C). Thus, it appears that in biological neural networks economical wiring and tight integration of local components are counterbalanced by the global minimization of processing steps across the network (but compare [23]). Indeed, when evaluated on a comparative scale for minimal wiring and minimal path lengths, the neural networks considered here were placed farther away from the optimal configuration for minimal wiring than from the best network arrangement for minimizing processing steps (Figure 5).

The importance of network shortcuts for reducing the number of processing steps has been pointed out before (e.g., [8]), particularly in the context of small-world network architectures [19]. The present study adds a spatial perspective to the previous topological investigations, by demonstrating that network shortcuts are formed mainly by long-distance connections. This conclusion, while intuitive, is not trivial, as one could also imagine alternative scenarios in which network shortcuts arise from short-distance connections (see Figure S3). The coincidence of long-distance connections with network shortcuts hints at a close match between the spatial layout and topology of neural networks [20]. It will be an interesting task for future studies to explore more fully the developmental and evolutionary reasons for this coincidence.

Minimizing average path length—that is, reducing the number of intermediate transmission steps in neural integration pathways—has several functional advantages. First, the number of intermediate nodes that may introduce interfering signals and noise is limited. Second, by reducing transmission delays from intermediate connections, the speed of signal processing and, ultimately, behavioural decisions is increased. Third, long-distance connections enable neighbouring as well as distant regions to receive activation nearly simultaneously [11,24] and thus facilitate synchronous information processing in the system (compare [25,26]). Fourth, the structural and functional robustness of neural systems increases when processing pathways (chains of nodes) are shorter. Each further node introduces an additional probability that the signal is not transmitted, which may be substantial (e.g., failure rates for transmitter release in individual synapses are between 50% and 90% [2]). Even when the signal survives, longer chains of transmission may lead to an increased loss of information. A similar conclusion, on computational grounds, was first drawn by John von Neumann [27] when he compared the organization of computers and brains. He argued that, due to the low precision of individual processing steps in the brain, the number of steps leading to the result of a calculation ("logical depth") should be reduced and highly parallel computing would be necessary. A loss of long-distance connections might also underlie pathological changes in neural network function. For instance, functional path length is increased in patients with Alzheimer's disease due to the loss of long-distance neural projections [28].

The advantages resulting from a short average number of processing steps appear as least as beneficial for the organization and function of neural networks as those commonly cited for wiring minimization. Previous analyses of cortical organization have demonstrated that the convoluted, laminar architecture of the mammalian cerebral cortex and the segregation into gray and white matter reduce total white matter volume and shorten projection lengths [29,30], also reducing conduction delays [31]. Ultimately, these adaptations point in the same direction as reductions in the number of processing steps, toward maximizing information-processing speed. Therefore, it is plausible that neural systems are adapted to more than just one design constraint, and that their observed organization is the outcome of an optimization of multiple parameters, which may be partly opposed to each other. For cortical networks, for example, additional constraints may arise from spatial factors that limit growth [11,32] or from critical periods for the establishment of cortical areas and their interconnections [33]. In addition to current ontogenetic constraints, the evolutionary history of a neural network might also conserve features of its predecessors, some of which may not be optimal for the present system [8].

In conclusion, we demonstrated that the organization of neural systems at different levels is not primarily shaped by a drive for minimal wiring. Rather, wiring minimization appears to be just one constraint among a variety of desirable factors [11,34] that also include the minimization of processing steps. It remains to be seen how, exactly, different structural, functional, and evolutionary constraints interact and compete to shape the organization of neural systems, and under which circumstances one singular constraint may dominate.

## Materials and Methods

**Primate corticocortical network.** We analyzed the spatial arrangement of 2,402 projections among 95 cortical areas and sub-areas of the primate (Macaque) brain. The connectivity data were retrieved from CoCoMac (http://www.cocomac.org [35]) and are based on three extensive neuroanatomical compilations [18,17,36] that collectively cover large parts of the cerebral cortex. Spatial positions of cortical areas were estimated from surface parcelling using the CARET software (http://brainmap.wustl.edu/caret). The spatial positions of areas were calculated as the average surface coordinate (or centre of mass) of the three-dimensional extension of an area (compare [11]). While the current cortical dataset is more extensive than those used in previous studies, it may still be partially incomplete, particularly for connections of motor, auditory, and somatosensory areas. The restriction arose from the fact that only studies could be used for which a parcellation scheme with spatial coordinates existed in CARET.

***C. elegans* neuronal networks.** We further analyzed two-dimensional spatial representations of the global neuronal network (277 neurons and 2,105 connections) of the nematode *C. elegans*, as well as a local subnetwork of neurons within *C. elegans* rostral ganglia (anterior, dorsal, lateral, and ring, 131 neurons with 764 unidirectional connections). Spatial two-dimensional positions (in the lateral plane), representing the position of the soma of individual neurons in *C. elegans*, were provided by Y. Choe [37]. Neuronal connectivity was obtained from [38]. This compilation is largely based on the dataset of White et al. [39] in which connections were identified by electron microscope reconstructions. The previously presented connectivity data [38] were modified in the following way. Neurons of the pharyngeal ring, for which there was no internal connection information [38], were removed from the network, leaving 280 neurons. In addition, three neurons (AIBL, AIYL, and SMDVL) had to be removed, because their positions were not provided in the set of spatial coordinates. Eventually 277 neurons were included in the analyses. The size of the global and local *C. elegans* datasets analyzed here was comparable to that used in previous studies. For example, studies of the small-world properties [19] or characteristic motifs [40] of *C. elegans* considered 282 and 187 neurons, respectively. Both chemical and electric synapses (gap junctions) were included as connections in our analysis.

Wiring length was calculated as direct Euclidean distance between connected components in three dimensions (Macaque, compare [11]) or two dimensions (*C. elegans*). Both datasets are available at http://www.biological-networks.org.

**Component placement optimization analysis.** In line with the definition of CPO [3], we investigated the possibility that spatial permutations of network components would lead to reductions in total wiring length of the neural networks. Because an exhaustive search of all possible $95! = 10^{148}$ (primate) or $139!$ and $256!$ (*C. elegans*) component rearrangements was computationally unattainable, we employed a stochastic optimization algorithm to specifically search for wiring length reductions.

The network layouts were explored by simulated annealing [41], which is a popular algorithm for combinatorial optimization. We implemented a Matlab version (MathWorks, Natick, Massachusetts, United States) of a standard algorithm [42]. Briefly, at each step, a cyclic permutation of three randomly chosen node positions was performed, after which the procedure recalculated the total wiring length of the networks. During the initial steps of the algorithm, decreases as well as increases of wiring length were carried through to the next stage of the optimization, while during later steps the selection of solutions was more strongly biased toward decreases in total wiring length. By this mechanism, the procedure could escape local minima and approximated solutions close to or at the global minimum. The wiring length for primate as well as *C. elegans* networks converged to a minimum after 40–60 and 8–12 steps, respectively (Figure S1). The simulated annealing process was performed independently 50 times on the original placement configuration. The minimum wiring length out of all trials was then used as an indicator of the possible reduction in total fibre length.

In addition, we exhaustively tested all possible two- and three-node permutations of the original network component arrangements, to establish the proportion of alternative component arrangements possessing a lower total wiring length. The total numbers of arrangements tested in the primate, local *C. elegans*, and global *C. elegans* were 8,930, 17,556, and 76,452 configurations for two-node permutations and 830,490, 2,299,836, and 21,024,300 configurations for three-node permutations, respectively.

Similar simulating annealing approaches were employed to create benchmark networks in which components were rearranged for maximum total wiring length, or in which connections were rewired to minimize or maximize average path lengths across the network. In each case, 20 runs of simulated annealing were performed and the minimal or maximal values were chosen as lower or upper limit for the benchmark, respectively.

**Minimal rewiring of networks.** To investigate the role of long-distance connections in neural systems, we also compared the original networks with networks of the same size (that is, with an identical number of nodes and connections) that had been rewired with the shortest possible connections. In order to generate such minimally rewired networks, possessing a greatly reduced proportion of long-distance connections, we employed the following procedure.

Starting with the spatial configuration of nodes in the original networks, but without edges, a minimum spanning tree was generated, to ensure that the resulting network would be connected. The minimum spanning tree for $N$ nodes consists of $N-1$ edges such that all nodes are part of the network and the total wiring length of all edges is minimal (compare [43]). The application of this initial step was required, since wiring of only the shortest available distances in *C. elegans* produced a fragmented network with multiple compartments. For the more densely connected cortical network of the primate, in contrast, the wiring of only the shortest distances already resulted in a connected network. However, for consistency, we started by creating the minimum spanning tree for both *C. elegans* and primate cortical networks. This constituted a conservative approach with regard to the subsequent computation of average path lengths.

In the next step, all pairwise distances of nodes were calculated and sorted by length. Starting with the shortest distance between any two nodes, edges between these nodes were generated until the total number of edges matched those of the original *C. elegans* or cortical networks. Thus, the resulting minimally rewired networks represented a lower bound for the wiring length of a connected network with the same total number of edges and identical node positions as in the original neural networks.

**Network characterization measures.** The following measures were used to characterize original and rearranged networks. Total wiring length was the sum of the metric length of all individual connections. In addition, total wiring volume included information about the anatomical strength of connections (as a first approximation of fibre diameter), as derived from tract-tracing experiments. The volume of an individual fibre was calculated as its metric length multiplied by its squared anatomical strength or density (as given by ordinal values 1, 2, or 3; 1 being the sparsest), with the total wiring volume as the sum of the volume of all individual fibres.

The average path length was the average number of connections that had to be passed on the shortest paths between all pairs of network nodes. It was calculated using Floyd's algorithm (compare [43]). The average metric path length described the metric distance that had to be travelled on average along the edges of the shortest path between any two network nodes. Therefore, it represented the average total wiring length of the shortest paths of the network. The clustering coefficient [19] was the proportion of actually present connections, out of all possible connections, among network nodes directly connected to a target node (i.e., the index measured the local connectivity among the target node's neighbours). The coefficient was calculated as the average over all individual nodes of the network.

**Control calculations for *C. elegans* network: Variations in connection site.** One possible confound of the *C. elegans* analyses was that exact positions of synapses were not included in the data, so morphological differences between neurons may have influenced the actual length of projections between neurons. Therefore, we tested the effect of variations in the position of synapses, by varying the connection distances between individual neurons through shifts in soma positions. As the animal extends mainly in the horizontal direction, variations in the spatial positions of synapses and cell bodies matter most along the horizontal axis. The position of each neuron along the longitudinal axis was shifted randomly rostrally or caudally, following a normal distribution with the mean around the actual position and a standard deviation of 10% of the total length of the worm (0.12 mm). Twenty different networks with varied longitudinal position were tested for potential reductions of total wiring length. A reduction of total wiring length was found that was almost as large as for the original positions (Figure S2). Therefore, random variations in neuronal positions did not alter the main finding.

However, could systematic errors, rather than random variations, substantially bias the results? The main sources of unknown synaptic positions are synapses along muscles and sensory endings. We noted that only 118 connections onto muscles are present, forming less than 6% of the total connections in *C. elegans* [38]. In the worst-case scenario, systematically overestimating the length of such connections might lead to an overestimation of the potential reduction in the real network. However, additional information about the detailed connection patterns of sensory and motor projections is unlikely to reduce the possible wire saving from 49% to 0%.

**Control calculations for *C. elegans* network: Variations of the third spatial coordinate.** Due to limitations of the currently available data, the *C. elegans* analyses involved only two dimensions. However, the third dimension, which forms the transversal axis, may not contribute much variation to the spatial positions of neurons. Because the animal extends mainly along one axis, it has been noted that the "layout problem is roughly one-dimensional" as "the length:diameter ratio of the worm body is about 20:1" [3]. In any case, we explored the effect of coordinate variations in the third dimension. Based on the used spatial database [37], neurons were classified as (a) placed on the left or right side of the animal, or (b) unspecified in their lateral position. In the simulation of a worst-case scenario, neurons of type (a) were positioned in the transversal axis as far away from the midline as possible; thus, neurons were placed at a distance of 57.3 μm from the midline, which represents the radius of the roundworm. Neurons of type (b) were placed on the origin of the z-coordinate (i.e., at the midline). The simulation of these conservative three-dimensional coordinates created a wiring length of 573.6 mm (only 7.8% more than for the original data in two dimensions). The application of the simulated annealing approach for reducing total wiring length then led to a total wiring length of 321.6 mm—that is, a reduction of 44%. Therefore, using three instead of two coordinates for *C. elegans* neuronal positions only slightly increased total wiring length, and did not change the potential for wiring length reduction.

**Control calculations for *C. elegans* network: Variations of synapse type.** The analyzed dataset of 277 nodes contained chemical as well as electrical (gap junction) synapses. Potentially, neurons connected by gap junctions may be more closely associated than neurons linked by chemical synapses, and the network layout may be affected by the type of interconnection. Therefore, we tested the effect of rearranging networks where electrical synapses were either present or absent. We used a smaller dataset of 256 neurons that included information on the type of synapse for each connection between neurons. We tested component rearrangement by simulated annealing with either chemical and electrical synapses or only chemical synapses present. In both cases a substantial reduction in total wiring length was found after rearrangement (for chemical and electrical synapses, 58% reduction on the global and 47% reduction on the local level; for chemical synapses only, 64% reduction on the global and 46% on the local level).

## Supporting Information

**Figure S1.** Successive Reduction of Total Wiring Length during Simulated Annealing Optimization

The curves summarize 50 individual trials for the Macaque cortical network (A) and the local *C. elegans* network (within rostral ganglia) (B). For each network, 50 simulated annealing runs were performed. All runs converged to a solution with shorter total wiring length than the original solution. For subsequent analysis, the single best configuration out of the 50 trials was used as the closest approximate of an optimal component placement solution.

Found at DOI: 10.1371/journal.pcbi.0020095.sg001 (258 KB TIF).

**Figure S2.** Random Variations in the Longitudinal Connection Site of Neurons in the *C. elegans* Neuronal Network

Dots represent the outcome of wiring optimization for 20 networks with randomly varied neuronal connection sites. The upper boundary of the diagram represents the total wiring length of the original network; the lower boundary represents the wiring optimization outcome for the original network.

Found at DOI: 10.1371/journal.pcbi.0020095.sg002 (100 KB TIF).

**Figure S3.** Linking Clusters by Short- or Long-Distance Connections

The two clusters are indicated by node colour. Note that both networks contain long-distance connections as well as the same number of connections, yet the link between the clusters (grey line) is alternatively provided by a short-distance connection (A) or a long-distance connection (B). The existence of long-distance connections, together with their effect on reducing the path length, suggests that the latter scenario does occur in neural systems.

Found at DOI: 10.1371/journal.pcbi.0020095.sg003 (165 KB TIF).

## Acknowledgments

We thank Yoonsuck Choe for providing us with spatial position data of *C. elegans* neurons and Veronica Miller for helpful comments on the manuscript.

**Author contributions.** MK and CCH conceived and designed the experiments. MK performed the experiments. MK analyzed the data. MK and CCH wrote the paper.

**Funding.** This work was supported by a fellowship from the German National Academic Foundation (M.K.).

**Competing interests.** The authors have declared that no competing interests exist.

### References

1. Laughlin SB, de Ruyter Van Steveninck RR, Anderson JC (1998) The metabolic cost of neural information. Nat Neurosci 1: 36–41.
2. Laughlin SB, Sejnowski TJ (2003) Communication in neuronal networks. Science 301: 1870–1874.
3. Cherniak C (1994) Component placement optimization in the brain. J Neurosci 14: 2418–2427.
4. Cherniak C, Changizi M, Kang DW (1999) Large-scale optimization of neuron arbors. Phys Rev E 59: 6001–6009.
5. Chklovskii DB, Schikorski T, Stevens CF (2002) Wiring optimization in cortical circuits. Neuron 34: 341–347.
6. Klyachko VA, Stevens CF (2003) Connectivity optimization and the positioning of cortical areas. Proc Natl Acad Sci U S A 100: 7937–7941.
7. Buzsaki G, Geisler C, Henze DA, Wang XJ (2004) Interneuron diversity series: Circuit complexity and axon wiring economy of cortical interneurons. Trends Neurosci 27: 186–193.
8. Striedter GF (2004) Principles of brain evolution. Sunderland (Massachusetts): Sinauer. pp. 217–253.
9. Braitenberg V, Schüz A (1998) Cortex: Statistics and geometry of neuronal connectivity. 2nd Ed. Springer. pp. 83–87, 129–134.
10. Hellwig B (2000) A quantitative analysis of the local connectivity between pyramidal neurons in layers 2/3 of the rat visual cortex. Biol Cybern 82: 111–121.
11. Kaiser M, Hilgetag CC (2004) Modelling the development of cortical networks. Neurocomp 58–60: 297–302.
12. Barbas H, Hilgetag CC, Saha S, Dermon CR, Suski JL (2005) Parallel organization of contralateral and ipsilateral prefrontal cortical projections in the rhesus monkey. BMC Neurosci 6: 32.
13. Cherniak C, Mokhtarzada Z, Rodriguez-Esteban R, Changizi K (2004) Global optimization of cerebral cortex layout. Proc Natl Acad Sci U S A 101: 1081–1086.
14. Chklovskii DB, Koulakov AA (2004) Maps in the brain: What can we learn from them? Annu Rev Neurosci 27: 369–392.
15. Cherniak C (1992) Local optimization of neuron arbors. Biol Cybern 66: 503–510.
16. Chklovskii DB (2004) Exact solution for the optimal neuronal layout problem. Neural Comput 16: 2067–2078.
17. Felleman DJ, van Essen DC (1991) Distributed hierarchical processing in the primate cerebral cortex. Cereb Cortex 1: 1–47.
18. Carmichael ST, Price JL (1994) Architectonic subdivision of the orbital and medial prefrontal cortex in the macaque monkey. J Comp Neurol 346: 366–402.
19. Watts DJ, Strogatz SH (1998) Collective dynamics of "small-world" networks. Nature 393: 440–442.
20. Young MP, Scannell JW (1996) Component placement optimization in the brain. Trends Neurosci 19: 413–415.
21. Ahn YY, Jeong H, Kim BJ (2006) Wiring cost in the organization of a biological neuronal network. Physica A 367: 531–537.
22. Hilgetag C, Barbas H (2006) Role of mechanical factors in the morphology of the primate cerebral cortex. PLoS Comput Biol 2: e22. DOI: 10.1371/journal.pcbi.0020022
23. Karbowski J (2001) Optimal wiring principle and plateaus in the degree of separation for cortical neurons. Phys Rev Lett 86: 3674–3677.
24. Masuda N, Aihara K (2004) Global and local synchrony of coupled neurons in small-world networks. Biol Cybern 90: 302–309.
25. von der Malsburg C (1995) Binding in models of perception and brain function. Curr Opin Neurobiol 5: 520–526.

26. Konig P, Engel AK, Roelfsema PR, Singer W (1995) How precise is neuronal synchronization? Neural Comput 7: 469–485.
27. von Neumann J (2000) The computer and the brain. New Haven and London: Yale University Press. pp. 74–79.
28. Stam CJ, Jones BF, Nolte G, Breakspear M, Scheltens P (2006) Small-world networks and functional connectivity in Alzheimer's disease. Cereb Cortex. E-pub ahead of print. Available: http://cercor.oxfordjournals.org/cgi/content/abstract/bhj127v1. Accessed 22 June 2006.
29. Ruppin E, Schwartz EL, Yeshurun Y (1993) Examining the volume efficiency of the cortical architecture in a multi-processor network model. Biol Cybern 70: 89–94.
30. Murre JM, Sturdy DP (1995) The connectivity of the brain: Multi-level quantitative analysis. Biol Cybern 73: 529–545.
31. Wen Q, Chklovskii DB (2005) Segregation of the brain into gray and white matter: A design minimizing conduction delays. PLoS Comput Biol 1: e78. DOI: 10.1371/journal.pcbi.0010078
32. Kaiser M, Hilgetag CC (2004) Spatial growth of real-world networks. Phys Rev E Stat Nonlin Soft Matter Phys 69: 036103.
33. Rakic P (2002) Neurogenesis in adult primate neocortex: An evaluation of the evidence. Nature Rev Neurosci 3: 65–71.
34. Sporns O, Chialvo DR, Kaiser M, Hilgetag CC (2004) Organization, development and function of complex brain networks. Trends Cogn Sci 8: 418–425.
35. Kötter R (2004) Online retrieval, processing, and visualization of primate connectivity data from the CoCoMac database. Neuroinf 2: 127–144.
36. Lewis J, van Essen DC (2000) Architectonic parcellation of parieto-occipital cortex and interconnected cortical regions in the macaque monkey. J Comp Neurol 428: 79–111.
37. Choe Y, McCormick BH, Koh W (2004) Network connectivity analysis on the temporally augmented *C. elegans* web: A pilot study. Soc Neurosci Abstr 30: 921.9.
38. Achacoso TB, Yamamoto WS (1992) AY's Neuroanatomy of *C. elegans* for computation. Boca Raton: CRC Press. pp. 79–164.
39. White JG, Southgate E, Thomson JN, Brenner S (1986) The structure of the nervous system of the nematode *Caenorhabditis elegans*. Philos Trans R Soc Lond B Biol Sci 314: 1–340.
40. Reigl M, Alon U, Chklovskii DB (2004) Search for computational modules in the *C. elegans* brain. BMC Biology 2: 25.
41. Metropolis N, Rosenbluth AW, Rosenbluth MN, Teller AH, Teller E (1953) Equation of state calculations by fast computing machines. J Chem Phys 21: 1087–1092.
42. Press WH, Teukolsky SA, Vetterling WT, Flannery BP (1994) Numerical recipes in C. 2nd Ed. Cambridge: Cambridge University Press. pp. 444–455.
43. Cormen TH, Leiserson CE, Rivest RL, Stein C (2001) Introduction to algorithms. 2nd Ed. Cambridge (Massachusetts): MIT Press. pp. 561–566, 629–632.